\documentclass[graybox, envcountchap]{svmult}

\usepackage{mathptmx}        
\usepackage{amsmath}
\usepackage{amssymb}
\usepackage{color}
\usepackage{helvet}          
\usepackage{courier}         
\usepackage{dirtree}
\usepackage{bm}

\usepackage{makeidx}         
\usepackage{graphicx}        
\usepackage{subfig}

\usepackage{multicol}        
\usepackage[bottom]{footmisc}

\usepackage{hyperref}        
\hypersetup{colorlinks=true,urlcolor=blue}

\usepackage[misc]{ifsym}

\makeindex             

\newcommand{\ie}{i.e.,~}
\newcommand{\eg}{e.g.,~}

\usepackage{physics}

\begin{document}


\title{General-Relativistic Magnetohydrodynamic Equations: the bare essential}
\author{Yosuke Mizuno and Luciano Rezzolla}

\institute{
  Yosuke Mizuno \at Tsung-Dao Lee Institute, Shanghai Jiao Tong
  University, 520 Shengrong Road, Shanghai 201210, China\\
  School of Physics and Astronomy, Shanghai Jiao Tong University, 
  800 Dongchuan Road, Shanghai 200240, China
%
  \and
  Luciano Rezzolla \at Institute for Theoretical Physics, Goethe
  University Frankfurt, Max-von-Laue-Strasse 1, 60438 Frankfurt am Main,
  Germany\\
  Frankfurt Institute for Advanced Studies, 60438, Frankfurt am Main,
  Germany \\
  School of Mathematics Trinity College, Dublin 2, Ireland
}
%
%
\maketitle

\abstract{Recent years have seen a significant progress in the development
  of general relativistic codes for the numerical solution of the
  equations of magnetohydrodynamics in spacetimes with high and dynamical
  curvature. These codes are valuable tools to explore the large-scale
  plasma dynamics such as that takes place when two neutron stars collide
  or when matter accretes onto a supermassive black hole. This chapter is
  meant to provide a very brief but complete overview of the set of
  equations that are normally solved in modern numerical codes after they
  are cast into a conservative formulation within a 3+1 split of
  spacetime.}


\section{Introduction}

Relativistic astrophysics studies the most energetic and violent
astrophysical processes that are characterized by very high speeds,
strong gravitational fields, very high temperatures and ultra-intense
magnetic fields. Under these conditions, which are normally met near
neutron stars and black holes, a fully general relativistic treatment is
necessary for an accurate description of the physical conditions.

In this context, relativistic magnetohydrodynamics (MHD) represents a
very effective framework to describe the dynamics of macroscopic plasma
in a relativistic regime, also when considering non-astrophysical
scenarios, such as the collision of heavy ions (see, \eg~\cite{Roy2015,
  Pu2016b}). It is in fact important to note that plasma is the most
diffused state of matter in the Universe and that most plasmas are
electrically conducting. In the MHD approximation, the plasma is treated
as a macroscopic fluid that is coupled with electromagnetic fields that
it produces with its dynamics or that may be present from external
sources. In addition, the collision times between particles in
astrophysical fluids are usually smaller than the other relevant
timescales of the systems. As a result, the mean free paths of the plasma
constituents are shortened, causing particles to interact on much smaller
spatial scales than those of the underlying macroscopic system. As a
result, astrophysical plasmas represent what is referred to as
collisional systems so that, despite being composed of particles at a
microscopic level, the plasmas can be described as a continuous medium
with well-defined macroscopic average quantities such as velocity,
density, and pressure.

Relativistic MHD is often employed to study the dynamics of relativistic
plasma, such as the collision of two magnetized neutron stars and the
ensuing gamma-ray burst~\cite{Liu:2008xy, Anderson2008, Rezzolla:2011,
  Palenzuela2013a, Kiuchi2015, Murguia-Berthier2016, Ciolfi2020c,
  Nathanail2020b, Nathanail2020c, Gottlieb2022}, or the accretion and
outflows onto a central compact object~\cite{Koide00, DeVilliers03,
  Mizuno04, McKinney2006c, Tchekhovskoy2011, Narayan2012, Mizuno2018,
  Liska2018, White2020, Osorio2022}.  Under these conditions, general
relativistic effects become important if not dominant and, hence, the
solution of the full set of general-relativistic MHD (GRMHD) equations
represent the only avenue to obtain an accurate and physically consistent
description of the system.  Numerical simulations have proven to be
crucial in modern theoretical astrophysics, enhancing our understanding
of the dynamics of astrophysical systems in highly nonlinear
regimes. From the point of view of GRMHD, over the past few decades, many
GRMHD codes have been developed \cite{Hawley84a, Koide00, DeVilliers03a,
  Gammie03, Baiotti04, Duez05MHD0, Shibata05b, Anninos05c, Anderson2006a,
  Anton06, Mizuno06, DelZanna2007, Giacomazzo:2007ti, Cerda2008,
  Etienne2015, Zanotti2015, White2016, Porth2017, Olivares2019,
  Cipolletta:2019geh, Cheong2021, Liska2019, Begue2023} employing the 3+1
decomposition of spacetime and conservative ‘Godunov’ schemes based on
approximate Riemann solvers \cite{Giacomazzo:2005jy, Rezzolla2013,
  Font03, Marti2015}. These codes are utilized to study
various high-energy astrophysical phenomena. Some of these GRMHD codes
incorporate radiation \cite{Sadowski2013, McKinney2014, Takahashi2016},
and/or non-ideal MHD processes \cite{Bucciantini2012a,
  Dionysopoulou:2012, Dionysopoulou:2012pp, Ressler2015, Chandra2015,
  Chandra2017, Qian2016, DelZanna2018, Ripperda2019}.  State-of-the-art
GRMHD codes implement full adaptive mesh refinement \cite{Porth2017,
  Olivares2019, Zanotti2015, Stone2020, Liska2019}, which is useful for
obtaining higher spatial resolution in particularly interesting regions
such as strong shocks, turbulence, and shear regions.

Chapter 1 of this book provides an overview of some essential properties
of GRMHD equations. The structure of the chapter is as follows.
Sec.~\ref{sec:1.1} introduces the covariant GRMHD equations, while
Sec.~\ref{sec:3+1} introduces the basic decomposition of a
four-dimensional spacetime into a timelike time-line and space-like
hypersurface. Section~\ref{sec:1.4} finally provides the form of the
GRMHD equations when cast in a conservative form within a 3+1 split
spacetime. A brief summary is contained in Sec.~\ref{sec:1.5}.

Before concluding, two important remarks should be made. First, we will
not discuss here the numerical methods that are normally employed to
solve the equations of GRMHD. This is because they are often complex,
involving a variety of Riemann solvers and reconstruction methods, and
hence needing an independent discussion; the interested reader can find
an introduction to such methods in a number of textbooks (see,
\eg~\cite{Leveque2002,Toro09, Rezzolla_book:2013, Marti2015,
  Balsara2017}) and more code-specific details in Part II of this book.
Second, in this chapter, we will not discuss the solution to the Einstein
equations, which are needed to account for the evolution of spacetime and
will be covered in Chapter 2. Throughout the text, we adopt units where the
speed of light, $c = 1$, and the gravitational constant, $G = 1$ and we
absorb a factor of $\sqrt{4 \pi}$ of the magnetic field four-vector,
$b^\mu$. We also use the index notation for contracted indices, employ
Greek (Latin) letters for indices running between 0 (1) and 3, and a
metric with signature $(-,+,+,+)$.

\section{Covariant General-Relativistic Magnetohydrodynamic equations}
\label{sec:1.1}

Hereafter, we will generally consider an ideal fluid endowed with
electromagnetic fields whose equations of motion, -- that is, the
general-relativistic magnetohydrodynamic (GRMHD) equations -- can be
derived after imposing the local conservation laws of rest-mass (the
continuity equation) and of the energy-momentum tensor, $T_{\mu \nu}$
(the Bianchi identities):
\begin{eqnarray}
  \label{eq:1-1}
  &&\nabla_\mu \tilde{J}^\mu = \nabla_\mu \rho u^\mu = 0\,,\\
  \nonumber \\
  \label{eq:1-2}
  &&\nabla_\mu T^{\mu \nu} = 0\,,
\end{eqnarray}
where $\nabla_\mu$ is the covariant derivative associated with the
four-dimensional spacetime metric $g_{\mu \nu}$, $\tilde{J}^{\mu}$ is the
rest-mass density current and $T^{\mu \nu}$ is the energy-momentum tensor
of the plasma.

Equation (\ref{eq:1-1}) represents the well-known mass-conservation law, where
$u^\mu$ is the fluid four-velocity and $\rho$ is the proper rest-mass
density. After introducing the projector operator orthogonal to $u_{\mu}$,
i.e.,
\begin{equation}
  h_{\mu\nu} := u_{\mu} u_{\nu} + g_{\mu\nu}\,,
\end{equation}
and such that $h_{\mu\nu} u^{\mu} = 0$, it is possible to realise that
Eqs. (\ref{eq:1-2}) are four distinct equations representing respectively
the conservation of energy
\begin{equation}
  \label{eq:en_cons}
  u_{\nu} \nabla_\mu T^{\mu \nu} = 0\,,
\end{equation}
and of four-momentum 
\begin{equation}
  \label{eq:mom_cons}
  h_{\alpha\nu} \nabla_\mu T^{\mu \nu} = 0\,.
\end{equation}

Note that the total energy-momentum is the linear combination of the
contributions coming from the matter and from the electromagnetic fields,
i.e., $T^{\mu \nu} := T^{\mu \nu}_{\rm m} + T^{\mu \nu}_{\rm f}$, where
\begin{equation}
T^{\mu \nu}_{\rm m} := \rho h u^\mu u^\nu + p g^{\mu \nu},
\end{equation}
and 
\begin{equation}
T^{\mu \nu}_{\rm f} := F^\mu_\lambda F^{\lambda \nu}- \frac{1}{4} (F^{\lambda
  \delta} F_{\lambda \delta}) g^{\mu \nu}\,.
\end{equation}
In the expressions above,
\begin{equation}
h:=1 + \epsilon + \frac{p}{\rho}\,,
\end{equation}
is the specific enthalpy, $\epsilon$ is specific internal energy, $p$ is
the fluid pressure, and $F^{\mu \nu}$ is the Faraday tensor.

The presence of electric and magnetic fields in the GRMHD equations
requires the solution of additional equations expressing the
corresponding conservation laws, namely, the Maxwell equations
\begin{equation}
  \nabla_\mu F^{\mu \nu} = \mathcal{J}^\mu\,,
  \label{eq:Maxwell1}
\end{equation}
\begin{equation}
  \nabla_\mu\, ^{*\!}F^{\mu \nu} = 0\,,
  \label{eq:Maxwell2}
\end{equation}
where $\mathcal{J}^\mu$ is the charge current density, and $^{*\!}F^{\mu
  \nu}$ is the dual of the Faraday tensor. The Faraday tensor $F^{\mu
  \nu}$ is constructed from the electric and magnetic fields, $E^\alpha$
and $B^\alpha$, as measured in the generic frame having $U^\alpha$ as
tangent vector, i.e.,
\begin{equation}
F^{\mu \nu} = U^\mu E^\nu - U^\nu E^\mu - \sqrt{-g} \eta^{\mu \nu \lambda
  \delta} U_{\lambda} B_{\delta}\,,
\end{equation}
where $\eta^{\mu \nu \lambda \delta}$ is the fully-antisymmetric symbol
(see, e.g., \cite{Rezzolla_book:2013}) and $g$ is the determinant of the
spacetime four-metric. The dual Faraday tensor
\begin{equation}
^*F^{\mu \nu} :=
\sqrt{-g} \eta^{\mu \nu \lambda \delta} F_{\lambda \delta} \,,
\end{equation}
is written as
\begin{equation}
^*F^{\mu \nu} = U^\mu B^\nu - U^\nu B^\mu + \sqrt{-g} \eta^{\mu \nu
    \lambda \delta} U_{\lambda} E_{\delta}\,.
\end{equation}

Most of the GRMHD simulations to date have explored scenarios within the
so-called ``ideal MHD limit'', that is, a limit in which the electrical
conductivity is assumed to be infinite. This limit represents a rather
good first approximation in astrophysical plasmas, where the conductivity
is actually very large\footnote{To be more precise, in geometrized units,
as the one adopted here, the induction equation reveals that the scalar
component of the electrical conductivity tensor $\sigma_{ij} = \sigma
\delta_{ij}$ can be expressed as the ratio between the Ohmic diffusion
timescale $\tau_{\rm diff}$ and the (square of the) dynamical timescale
$\tau_{\rm dyn}$ of the magnetic field, \ie $\sigma \simeq \tau_{\rm
  diff}/\tau^2_{\rm dyn}$~\cite{Harutyunyan2018}. In typical
astrophysical plasmas, the dynamical timescales of the system are much
shorter than the timescale associated with the diffusion of the the
magnetic field so that it is reasonable to assume that the conductivity
is actually infinite.}. Under these conditions, the electric charges are
``infinitely effective'' in canceling any electric field, which are
therefore zero in the frame comoving with the fluid $u^\mu$, \ie
\begin{equation}
  F^{\mu \nu} u_\nu = 0\,.
  \label{eq:MHDcondition}
\end{equation}
The main consequence of the condition~\eqref{eq:MHDcondition} is that the
electric fields cease to be independent vector fields and can be obtained
from simple algebraic expressions involving the fluid four-velocity and
the magnetic fields. In particular, after defining the electric and
magnetic four-vectors in the fluid frame as
\begin{eqnarray}
  e^\mu := F^{\mu \nu} u_\nu\,, \\
  b^\mu := ^*F^{\mu \nu} u_\nu\,,
\end{eqnarray}
with the constraints that 
\begin{equation}
e^\mu = 0\,,
\end{equation}
and that the comoving magnetic field is fully spatial
\begin{equation}
u_\mu b^\mu = 0\,.
\end{equation}
Under these conditions, the Faraday tensor can then be rewritten as
\begin{eqnarray}
 && F^{\mu \nu} = -\sqrt{-g} \eta^{\mu \nu \lambda \delta} u_\lambda b_\delta\,,\\
 && ^*F^{\mu \nu} = b^\mu u^\nu - b^\nu u^\mu\,.
  \label{eq:1-11}
\end{eqnarray}

We can write the total energy-momentum tensor in terms of the vectors
$u^\mu$ and $b^\mu$ \cite{Anile1990} as
\begin{equation}
T^{\mu \nu} = \rho h_{\rm tot} u^\mu u^\nu + p_{\rm tot} g^{\mu \nu} - b^\mu b^\nu\,,
\end{equation}
where we introduced total pressure
\begin{equation}
p_{\rm tot} := p + \frac{b^2}{2}\,,
\end{equation}
which now includes the magnetic pressure
\begin{equation}
p_{\rm mag} := \frac{1}{2}b^{\mu}b_{\mu} =: \frac{1}{2}b^2\,,
\end{equation}
while the total specific enthalpy is given by
\begin{equation}
h_{\rm tot} := h + \frac{b^2}{\rho}\,.
\end{equation}
Note that the square of the magnetic-field strength in the fluid frame
also satisfies the following identity $b^2 = B^2 - E^2 > 0$, which is
sometimes used as a physical constraint in numerical
evolutions~\cite{Palenzuela:2010b, Alic:2012}. Finally, it is useful to
express the current density $\mathcal{J}^\mu$ into two components, \ie a
``convection'' (or advection) term and a ``conduction'' term
\begin{equation}
  \label{eq:curr_dens}
  \mathcal{J}^\mu = q n^\mu + J^\mu,
\end{equation}
where $q= - \mathcal{J}^\mu n_\mu$ is the charge density, $q n^{\mu}$ is
the convection current and $J^\mu$ is conduction current, that is, the
current density measured by the Eulerian observer and such that $J^\mu
n_\mu = 0$. 
  
In summary, the set of covariant GRMHD equations consists of the coupled
system of two conversation laws \eqref{eq:1-1}--\eqref{eq:1-2} and of the
two sets of Maxwell equations \eqref{eq:Maxwell1} and
\eqref{eq:Maxwell2}. However, as such, the corresponding system of
equations is not closed and needs to be complemented with an equation of
state that normally prescribes the behavior of the pressure as a
function of the rest-mass density, specific internal energy (temperature)
and particle abundances. The complexity of such equations of state varies
enormously, from very simple and analytic ones -- as those employed in
simulations of accretion onto supermassive black holes -- to very
sophisticated and tabulated ones -- as those employed in simulations of
binary neutron stars. Chapter 3 of this book will be dedicated to a
detailed discussion of the equations of state employed in modern
numerical simulations.

However, more importantly, as presented here,
Eqs.~\eqref{eq:1-1}--\eqref{eq:1-2} and
\eqref{eq:Maxwell1}--\eqref{eq:Maxwell2} are not particularly useful for
a numerical solution in a simulation code; rather, they first need to be
cast within a 3+1 decomposition of spacetime and then expressed in a
conservative formulation, as we will discuss in the following two
sections.

\section{The 3+1 decomposition of spacetime}
\label{sec:3+1}

The intrinsically ``covariant view'' of Einstein's theory of general
relativity is based on the concept that all coordinates are equivalent
and, hence, the distinction between spatial and time coordinates is more
an organizational matter than a strict requirement of the theory. Yet,
our experience, and the laws of physics on sufficiently large scales, do
suggest that a distinction between the time coordinate from the spatial
ones is the most natural one in describing physical
processes. Furthermore, such a distinction of time and space is the
simplest way to exploit a large literature on the numerical solution of
hyperbolic partial differential equations as those of relativistic
MHD. Adopting this principle, and following closely the presentation
already offered in Ref.~\cite{Rezzolla_book:2013}, we ``foliate''
spacetime in terms of a set of non-intersecting spacelike hypersurfaces
$\Sigma := \Sigma(t)$, each of which is parameterized by a constant value
of the coordinate $t$. In this way, the three spatial coordinates are
split from the one temporal coordinate and the resulting construction is
called the {3+1 decomposition} of spacetime~\cite{MTW1973}.

\begin{figure}
\begin{center}
\includegraphics[angle=0,width=11.0cm]{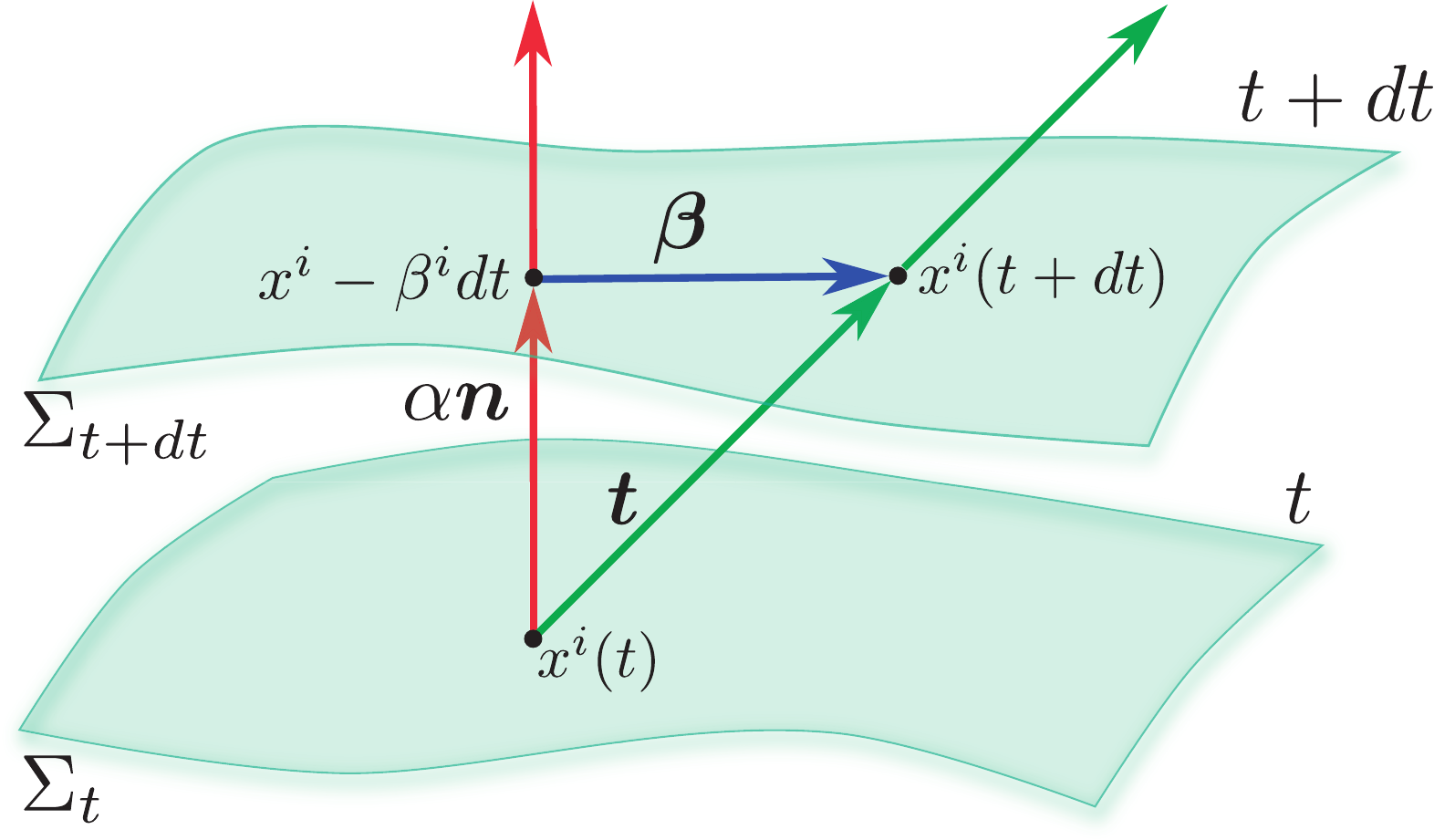}
\caption{Schematic representation of the 3+1 decomposition of spacetime
  with hypersurfaces of constant time coordinate $\Sigma_t$ and
  $\Sigma_{t+dt}$ foliating the spacetime. The four-vector $\bm{t}$
  represents the direction of evolution of the time coordinate $t$ and
  can be split into a timelike component $\alpha\bm{n}$, where $\bm{n}$
  is a timelike unit normal to the hypersurface, and into a spacelike
  component, represented by the spacelike four-vector $\bm{\beta}$. The
  function $\alpha$ is the ``lapse'' and measures the proper time between
  adjacent hypersurfaces, while the components of the ``shift'' vector
  $\beta^{i}$ measure the change of coordinates from one hypersurface to
  the subsequent one. Figure reproduced
  from~\protect{\cite{Rezzolla_book:2013}}.}
\label{fig:3p1}
\end{center}
\end{figure}

Given one such constant-time hypersurface, $\Sigma_t$, belonging to the
{foliation} $\Sigma$, we can introduce a timelike four-vector
$\bm{n}$ normal to the hypersurface at each event in the spacetime and
such that its dual one-form $\bm{\Omega} := \bm{\nabla} t$ is
parallel to the gradient of the coordinate $t$, \ie
\begin{equation}
\label{eq:n_mu}
n_\mu=A \Omega_{\mu} = A \nabla_\mu t  \,,
\end{equation}
with $n_{\mu} = \{A,0,0,0 \}$ and $A$ a constant to be determined. If we
now require that the four-vector $\bm{n}$ defines an observer and thus
that it measures the corresponding four-velocity, then from the
normalization condition on timelike four-vectors, $n^\mu n_\mu=-1$, we
find that
\begin{equation}
\label{eq:nmu_nnu}
n^\mu n_\mu=g^{\mu\nu}n_\mu n_\nu = 
g^{tt}A^2=-\frac{1}{\alpha^2}A^2=-1 \,,
\end{equation}
where we have defined $\alpha^2 := - 1/g^{tt}$. From the last equality in
expression \eqref{eq:nmu_nnu} it follows that $A=\pm\alpha$ and we will
select $A=-\alpha$, such that the associated vector field $n^\mu$ is
future directed. The quantity $\alpha$ is commonly referred to as
the \emph{lapse} function, it measures the rate of change of the
coordinate time along the vector $n^\mu$ (see Fig.~\ref{fig:3p1}), and
will be a building block of the metric in a 3+1 decomposition [cf.,
Eq. \eqref{eq:ds2_3p1}].

The specification of the normal vector $\bm{n}$ allows us to define the
metric associated to each hypersurface, \ie
\begin{align}
&\gamma_{\mu \nu} := g_{\mu \nu} + n_{\mu} n_{\nu}\,,
&\gamma^{\mu \nu} := g^{\mu \nu} + n^{\mu} n^{\nu}\,,
\end{align}
where $\gamma^{0\mu}=0$, $\gamma_{ij} = g_{ij}$, but in general
$\gamma^{ij} \neq g^{ij}$. Also note that $\gamma^{ik} \gamma_{kj}=
\delta^i_j$, that is, $\gamma^{ij}$ and $\gamma_{ij}$ are the inverse of
each other, so that the spatial metric $\bm{\gamma}$ can be used for
raising and lowering the indices of purely spatial vectors and tensors.

The tensors $\bm{n}$ and $\bm{\gamma}$ provide us with two useful tools
to decompose any four-dimensional tensor into a purely spatial part
(hence contained in the hypersurface $\Sigma_t$) and a purely timelike
part (hence orthogonal to $\Sigma_t$ and aligned with $\bm{n}$). Not
surprisingly, the spatial part is readily obtained after contracting with
the \emph{spatial projection operator} (or \emph{spatial projection
  tensor})
\begin{equation}
\label{eq:space_proj}
\gamma^{\mu}_{\ \,\nu} := g^{\mu \alpha} \gamma_{\alpha \nu} =
g^{\mu}_{\ \,\nu} + n^{\mu} n_{\nu} = \delta^{\mu}_{\ \,\nu} + n^{\mu} n_{\nu}\,,
\end{equation}
while the timelike part is obtained after contracting with the \emph{time
  projection operator} (or \emph{time projection tensor})
\begin{equation}
\label{eq:time_proj}
N^{\mu}_{\ \,\nu} := - n^{\mu} n_{\nu}\,,
\end{equation}
and where the two projectors are obviously orthogonal, \ie
\begin{equation}
\gamma^{\alpha}_{\ \,\mu} N^{\mu}_{\ \nu} = 0\,.
\end{equation}
Hence, a generic four-vector $\bm{U}$ can be decomposed as
\begin{equation}
U^\mu=\gamma^\mu_{\ \,\nu} U^\nu + N^\mu_{\ \nu} U^\nu\,,
\end{equation}
where the \emph{purely spatial} part $\gamma^\mu_{\ \,\nu} U^\nu=V^\mu$
is still a four-vector that, by construction, has a zero contravariant
time component, \ie $V^t=0$, whereas it has the covariant time component,
$V_t = g_{\mu t}V^\mu$, which is nonzero in general. Analogous
considerations can be done about tensors of any rank.

We have already seen in Eq. \eqref{eq:n_mu} that the unit normal
$\bm{n}$ to a spacelike hypersurface $\Sigma_t$ does not represent
the direction along which the time coordinate changes, that is, it is not
the direction of the time derivative. Indeed, if we compute the
contraction of the two tensors we obtain
\begin{equation}
\label{eq:tpo_6}
n^{\mu} \Omega_{\mu} = \frac{1}{A} n^{\mu} n_{\mu} =
\frac{1}{\alpha} \neq 1\,.
\end{equation}
We can therefore introduce a new vector, $\bm{t}$, along which to carry
out the time evolutions and that is dual to the surface one-form
$\bm{\Omega}$. Such a vector is just the time-coordinate basis vector and
is defined as the linear superposition of a purely temporal part
(parallel to $\bm{n}$) and of a purely spatial one (orthogonal to
$\bm{n}$), namely
\begin{equation}
\label{eq:tpo_7}
\bm{t} = \bm{e}_t = \partial_t := \alpha \bm{n} + \bm{\beta}\,.
\end{equation}
The purely spatial vector $\bm{\beta}$ [\ie $\beta^{\mu} = (0,\beta^i)$]
is usually referred to as the {shift vector} and will be another building
block of the metric in a 3+1 decomposition [cf.,
  Eq. \eqref{eq:ds2_3p1}]. The decomposition of the vector $\bm{t}$ into
a timelike component $\alpha\bm{n}$ and a spatial component $\bm{\beta}$
is shown in Fig.~\ref{fig:3p1} (note that $\alpha=1, \beta^i=0$ in
special relativity).

We can check that $\bm{t}$ is a coordinate basis vector by verifying that
\begin{equation}
\label{eq:tpo_8}
t^{\mu} \Omega_{\mu} = 
\alpha n^{\mu} \Omega_{\mu} + \beta^{\mu} \Omega_{\mu} = 
\frac{\alpha}{\alpha} = 1\,,
\end{equation}
from which it follows that the vector $\bm{t}$ is effectively
dual to the one-form $\bm{\Omega}$. This guarantees that the
integral curves of $t^{\mu}$ are naturally parameterized by the time
coordinate. As a result, all infinitesimal vectors $t^{\mu}$ originating
on one hypersurface $\Sigma_t$ would end up on the same hypersurface
$\Sigma_{t+dt}$. Note that this is not guaranteed for translations along
$\Omega_{\mu}$ and that since $t^{\mu}t_{\mu} = g_{tt}= -\alpha^2 +
\beta^{\mu}\beta_{\mu}$, the vector $\bm{t}$ is not necessarily timelike
(the shift can in fact be superluminal).

Using the components of $\bm{n}$ 
\begin{align}
\label{eq:eulerian_observer}
&  n_\mu= \left(-\alpha,0,0,0 \right)\,, 
&& n^\mu=\frac{1}{\alpha}\left(1,-\beta^i \right)\,,
\end{align}
we can now express the generic \emph{line element} in a 3+1 decomposition
as
\begin{equation}
\label{eq:ds2_3p1}
ds^{2} = -(\alpha^{2}-\beta_{i}\beta^{i}) dt^{2}+ 
2 \beta_{i} dx^{i} dt + \gamma_{ij} dx^{i}dx^{j} \,.
\end{equation}
Expression \eqref{eq:ds2_3p1} clearly emphasises that when
$\beta^i=0=dx^i$, the lapse measures the proper time, $d\tau$, between
two adjacent hypersurfaces, \ie
\begin{equation}
\label{eq:lapse_pt}
d\tau^{2} = \alpha^{2}(t,x^j) dt^{2}  \,,
\end{equation}
while the shift vector measures the change of coordinates of a point from
the hypersurface $\Sigma_t$ to the hypersurface $\Sigma_{t+dt}$, \ie
\begin{equation}
x^i_{t+dt}=x^i_{t}-\beta^i (t,x^j) dt \,.
\end{equation}
Similarly, the covariant and contravariant components of the
metric \eqref{eq:ds2_3p1} can be written explicitly as
\begin{align}
&g_{\mu\nu} = \left(
\begin{array}{cc}
-\alpha^2 + \beta_i \beta^i ~~&~~ \beta_i   \\
~~&~~\\
 \beta_i            ~~&~~ \gamma_{ij} \\
\end{array} 
\right) \,,
&&g^{\mu\nu} = 
\left(
\begin{array}{cc}
-1/\alpha^2      ~~&~~ \beta^i/\alpha^2   \\
~~&~~\\
 \beta^i/\alpha^2   ~~&~~ \gamma^{ij} - \beta^i \beta^j/ \alpha^2 \\
\end{array} \right) \,,
\end{align}
from which it is easy to obtain an important identity which will be used
extensively hereafter, \ie
\begin{equation}
\label{eq:sqrtg}
\sqrt{-g} = \alpha \sqrt{\gamma}\,,
\end{equation}
where $g := \det(g_{\mu \nu})$ and $\gamma := \det(\gamma_{ij})$.

When defining the unit timelike normal $\bm{n}$ in
Eq. \eqref{eq:nmu_nnu}, we have mentioned that it can be associated to
the four-velocity of a special class of observers, which are referred to
as \emph{normal} or \emph{Eulerian observers}. Although this denomination
is somewhat confusing, since such observers are not at rest with respect
to infinity but have a coordinate velocity
\hbox{$dx^i/dt = n^i =-\beta^i/\alpha$}, we will adopt this traditional nomenclature 
also in the following and thus take an ``Eulerian observer''
as one with four-velocity given by (\ref{eq:eulerian_observer}).

When considering a fluid with four-velocity $\bm{u}$, the spatial
four-velocity $\bm{v}$ measured by an Eulerian observer will be given by
the ratio between the projection of $\bm{u}$ in the space orthogonal to
$\bm{n}$, \ie $\gamma^i_{\ \,\mu} u^{\mu} = u^i$, and the {Lorentz
  factor} of $\bm{u}$ as measured by $\bm{n}$~\cite{deFelice90}
\begin{equation}
-n_{\mu} u^{\mu} = \alpha u^t\,.
\end{equation}
As a result, the spatial four-velocity of a fluid as measured by an
Eulerian observer will be given by
\begin{align}
\label{projection_of_u_1}
\bm{v} := 
\frac{\bm{\gamma} \cdot \bm{u}}{-\bm{n}\cdot\bm{u}}\,,
\end{align}
or, in component form, by
\begin{align}
\label{projection_of_u_2}
&v^t = 0\,,&  v^i =
\frac{\gamma^i_{\ \,\mu} u^{\mu}}{\alpha u^t} = & \frac{1}{\alpha}
\left(\frac{u^i}{u^t} + \beta^i \right)\,,\\
\label{projection_of_u_3}
&v_t = \beta_i v^i  \,,& v_i =
\frac{\gamma_{i\mu} u^{\mu}}{\alpha u^t}=
\frac{u_i}{\alpha u^t} = &
\frac{\gamma_{ij}}{\alpha}
\left(\frac{u^j}{u^t} + \beta^j \right)\,.
\end{align}
Using now the normalisation condition $u^{\mu}u_{\mu}=-1$ and indicating
as usual with $W$ the {Lorentz factor}, we obtain
\begin{align}
\label{eq:LorFact}
& \alpha u^t = -\bm{n}\cdot\bm{u} =
\frac{1}{\sqrt{1 - v^iv_i}} = W\,,
&& u_t = W(-\alpha+\beta_i v^i) \,, 
\end{align}
so that the
components \eqref{projection_of_u_2}--\eqref{projection_of_u_3} can
finally, be written as
\begin{align}
\label{projection_of_u_4}
& v^i = \frac{u^i}{W} + \frac{\beta^i}{\alpha} = 
\frac{1}{\alpha}\left(\frac{u^i}{u^t} + \beta^i\right)\,,
&&v_i = \frac{u_i}{W} = \frac{u_i}{\alpha u^t} \,,
\end{align}
where in the last equality we have exploited the fact that
$\gamma_{ij}u^j = u_i - \beta_i W/\alpha$. Finally, using
expressions \eqref{projection_of_u_1} and \eqref{eq:LorFact}, it is also
possible to write the fluid four-velocity as
\begin{equation}
\label{split_of_u}
u^\mu=W(n^\mu + v^\mu)\,,
\end{equation}
which highlights the split of $\bm{u}$ into a temporal and a
spatial part.

\begin{figure}
\begin{center}
\includegraphics[angle=0,width=10.cm]{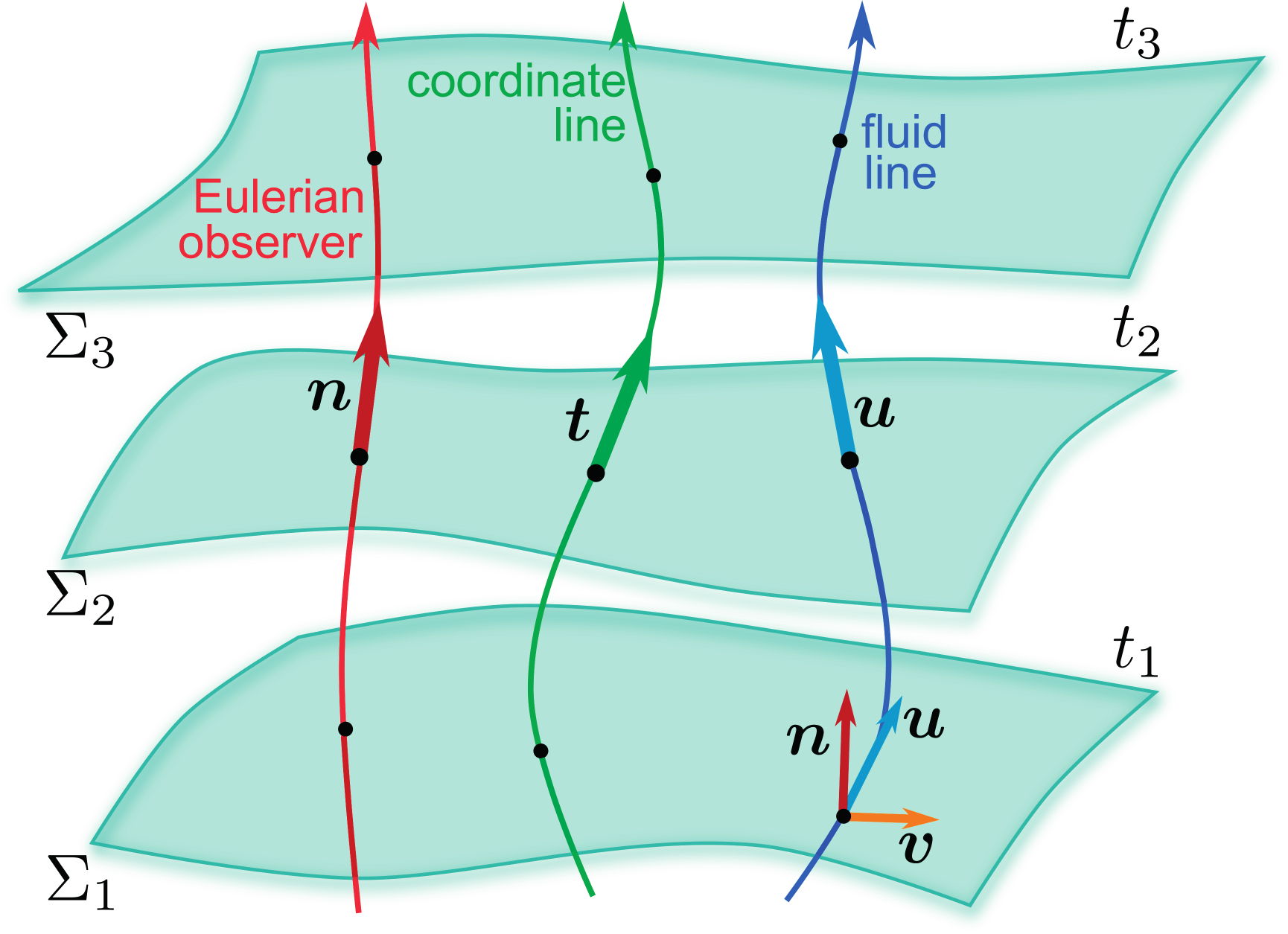}
\caption{Schematic representation of the different unit vectors in a 3+1
  decomposition of spacetime (see Fig.~\ref{fig:3p1}). The four-vectors
  $\bm{n}$, $\bm{t}$, and $\bm{u}$ represent the unit timelike normal,
  the time-coordinate basis vector and the fluid four-velocity,
  respectively. It is shown the associated worldlines, namely, the normal
  line, the coordinate line, and the fluidline. The spatial
  projection $\bm{v}$ of the fluid four-velocity $\bm{u}$ as measured by
  the (Eulerian) normal observer $\bm{n}$ are also shown. Figure reproduced
  from~\protect{\cite{Rezzolla_book:2013}}.}
\label{fig:unit_vectors}
\end{center}
\end{figure}

The three different unit four-vectors in a 3+1 decomposition of spacetime
are shown in Fig.~\ref{fig:unit_vectors}, which should be compared with
Fig.~\ref{fig:3p1}. The four-vectors $\bm{n}$, $\bm{t}$ and $\bm{u}$
represent the unit timelike normal, the time-coordinate basis vector and
the fluid four-velocity, respectively. Also shown are the associated
worldlines, namely, the {normal line} representing the worldline of an
Eulerian observer, the {coordinate line} representing the worldline of a
coordinate element, and the {fluidline}. The figure also reports the
spatial projection $\bm{v}$ of the fluid four-velocity $\bm{u}$, thus
highlighting that $\bm{v}$ is the three-velocity as measured by the
normal observer.

What remains to be done at this point is to apply the 3+1 decomposition
to the relevant vector fields that appear in the GRMHD equations. In
particular, we start by expressing the Faraday tensor and the dual of the
Faraday tensor in Maxwell's equations (\ref{eq:Maxwell1}) and
(\ref{eq:Maxwell2}) respectively as
\begin{align}
F^{\mu \nu} &= n^\mu E^\nu - n^\nu E^\mu - \sqrt{-g} \eta^{\mu \nu
  \lambda \delta} n_{\lambda} B_{\delta}\,, \\
^*F^{\mu \nu} &= n^\mu B^\nu
- n^\nu B^\mu + \sqrt{-g} \eta^{\mu \nu \lambda \delta} n_{\lambda}
E_{\delta}\,,
\end{align}
where the spatial components of the electric and magnetic fields measured by
the Eulerian observer are given as
\begin{equation}
  \label{eq:EB field}
  E^i := F^{i \nu} n_{\nu} = \alpha F^{it}\,, \qquad \qquad
  B^i := ^*F^{i \nu} n_{\nu} = \alpha ^*F^{it}\,. 
\end{equation}
We should note that our definition of the electric and magnetic fields
differs by a factor $\alpha$ from the corresponding definition used in
Refs.~\cite{Komissarov1999, Gammie03}.

Going back to the definition of the total current
density~\eqref{eq:curr_dens} and recalling that the conduction current is
purely spatial, \ie $J^{\mu}n_{\mu}=0$, we can express its spatial
components in terms of what is otherwise referred to as Ohm's law
\cite{Palenzuela:2008sf, Bucciantini2012a},
\begin{equation}
\label{eq:Ohm}
J^i = q v^i + \frac{W}{\eta} \left[ E^i + \frac{1}{\sqrt{\gamma}}
  \eta^{ijk} v_j B_k - (v_k E^k)v^i \right]\,,
\end{equation}
where $\eta$ is the resistivity and is the inverse of the scalar term of
the conductivity tensor, \ie $\eta :=1/\sigma$~\cite{Harutyunyan2018},
where $\eta_{ijk}$ is a Levi-Civita antisymmetric symbol, and where we
have ignored the Hall or dynamo terms for simplicity (see
Refs.~\cite{Bucciantini2012a, Palenzuela:2008sf} where these terms are
included in a generalized Ohm's law). Assuming now the ideal MHD
condition expressed by Eq.~(\ref{eq:MHDcondition}), we can obtain the
explicit and algebraic expression between the electric and magnetic fields
\begin{equation}
  \label{eq:ideal_condition}
  E^i = \sqrt{\gamma} \eta^{ijk} B_j v_k\,.
\end{equation}
This results, which coincides with the equivalent expression in Newtonian
MHD, underlines the passive role of dependent quantity for the electric
field in the ideal-MHD equations.

Finally, using Eq.~(\ref{eq:1-11}) together with (\ref{eq:EB field}), we
can obtain the transformation between magnetic four-vector field in the
fluid frame $b^\mu$ and the magnetic four-vector field in the Eulerian frame
$B^\mu$ as 
\begin{equation}
\label{eq:fluid-b}
  b^{t} = \frac{W}{\alpha}(B^i v_i)\,, \qquad \qquad
  b^i = \frac{1}{W}(B^i + \alpha b^t u^i)\,,
\end{equation}
which allows us to express the dual Faraday tensor (\ref{eq:1-11}) in
terms of the magnetic four-vector field in the Eulerian frame as
\begin{equation}
^*F^{\mu \nu} = \frac{1}{W}(B^\mu u^\nu - B^\nu u^\mu)\,,
\end{equation}
and calculate the scalar $b^2$ as
\begin{equation}
b^2 =\frac{B^2 + \alpha^2 (b^t)^2}{W^2} =\frac{B^2}{W^2} + (B^i v_i)^2\,,
\end{equation}
where $B^2 := B^i B_i$.

\section{Formulation of the GRMHD equations for numerical simulations}
\label{sec:1.4}

\subsection{Conservative Formulations}

The equations of (relativistic) hydrodynamics and MHD can be written in
the generic first-order-in-time form
\begin{equation}
\label{hyper_conserv_2bis}
\partial_t \bm{U} + \bm{A} \cdot \bm{\nabla}\bm{U} = \bm{S}\,,
\end{equation}
and the system above is said to be \emph{hyperbolic} if the matrix of
coefficients $\bm{A}$ is diagonalisable with a set of real eigenvalues,
or \emph{eigenspeeds}, $\lambda_1,\ldots,\lambda_N$ and a corresponding
set of $N$ linearly independent \emph{right eigenvectors} $\bm{R}^{(1)},
\ldots, \bm{R}^{(m)}$, such that $\bm{A} \bm{R}^{(i)} =
\lambda_i\bm{R}^{(i)}$, $\bm{\Lambda} := \bm{R}^{-1} \bm{A} \bm{R} = {\rm
  diag}(\lambda_1, \ldots, \lambda_N)$ is the diagonal matrix of
eigenvalues and $\bm{R}$ the matrix of right eigenvectors.
 
The most important property of \emph{hyperbolic equations} is that they
\emph{are well-posed}, hence suitable for numerical solution. Moreover,
if the matrix $\bm{A}(\bm{U})$ is the Jacobian of a \emph{flux vector}
$\bm{F}(\bm{U})$ with respect to the state vector $\bm{U}$, namely if
$\bm{A}(\bm{U}) := \partial\bm{F} / \partial\bm{U}$, then the homogeneous
version of the system \eqref{hyper_conserv_2bis} can be written in
\emph{conservative form} as
\begin{equation}
\label{hyper_conserv_3}
\partial_t \bm{U} + \bm{\nabla} \bm{F}(\bm{U}) = 0 \,,
\end{equation}
where $\bm{U}$ is therefore called the vector of \emph{conserved
variables}. With this definition in mind, we can now discuss two theorems
underlining the importance of a conservative formulation. The first one
loosely speaking states that: {conservative numerical schemes,} -- that
is, a numerical scheme based on the conservative formulation of the
equations -- {if convergent, do converge to the weak solution of the
  problem}~\cite{Lax60}. The second theorem states instead that:
\textit{non-conservative schemes, \ie schemes in which the equations are
  not written in the conservative form \eqref{hyper_conserv_3}, do not
  converge to the correct solution if a shock wave is present in the
  flow}~\cite{Hou94}. In other words, the two theorems above state that
if a conservative formulation \emph{is used}, then we are guaranteed that
the numerical solution will converge to the correct one, while if a
conservative formulation \emph{is not used}, we are guaranteed to
converge to the incorrect solution in the likely event in which the flow
develops a discontinuity.

\subsection{The 3+1 Valencia formulation(s)}

Given the importance of a conservative formulation to obtain a well-posed
system of equations, more than 30 years ago the group of Valencia started
to develop Eulerian 3+1 formulations of the relativistic hydrodynamic and
MHD equations written in conservative forms. Because of this, these
formulations are often referred to as the ``Valencia formulations''.  The
first step in this direction was taken by considering the equations of
special-relativistic hydrodynamics, which were cast into a conservative
formulation~\cite{Marti91} and solved in conjunction with High-Resolution
Shock-Capturing (HRSC) methods~(see, \eg~\cite{Rezzolla_book:2013} for an
introduction to HRSC methods). The second step was taken a few years
later, when the equations of genera-relativistic hydrodynamics were cast
into a conservative formulation~\cite{Banyuls97} and have been employed
in a variety of numerical simulations starting from Refs.~\cite{Font02c,
  Baiotti04}. A final step, which is very relevant to the content of this
chapter, was taken when the equations of GRMHD were cast in an Eulerian
conservative formulation~\cite{Anton06}.

In what follows, we detail the steps that are needed in order to cast the
covariant set ideal-GRMHD equations \eqref{eq:1-1}, \eqref{eq:1-2},
\eqref{eq:Maxwell1}, \eqref{eq:Maxwell2} into the following 3+1 and
Eulerian conservative form
\begin{equation}
\label{eq:compact-form}
\partial_t \left(\sqrt{\gamma} \bm{U}\right) +\partial_i
\left(\sqrt{\gamma} \bm{F}^i\right) = \sqrt{\gamma} \bm{S}\,,
\end{equation}
where $\bm{U}$ is the vector of conserved variables, $\bm{F}^i$ are the
flux vectors (or ``fluxes''), and $\bm{S}$ is the vector of source terms.

We can start from the covariant continuity equation \eqref{eq:1-1}, which
can be written as
\begin{align}
\nabla_\mu \left(\rho u^\mu\right) &= \frac{1}{\sqrt{-g}} \partial_\mu
\left(\sqrt{-g} \rho u^\mu\right) \notag \\ &= \frac{1}{\sqrt{-g}}
\left[\partial_t \left(\sqrt{-g} \rho u^t\right) + \partial_i
  \left(\sqrt{-g} \rho u^i\right)\right] = 0\,, \label{eq:continuity}
\end{align}
which, after introducing the conserved rest-mass density in the Eulerian frame,
\begin{equation}
D := \rho u^\mu n_\mu = \rho \alpha u^t = \rho W\,,
\end{equation}
can be rewritten in the conservative form
\begin{equation}
  \label{eq:continuity2}
  \partial_t \left(\sqrt{\gamma} D\right) + \partial_i
  \left[\sqrt{\gamma} D \left(\alpha v^i -\beta^i\right)\right] = 0\,.
\end{equation}
Equivalently, after defining the transport velocity $\mathcal{V}^i :=
\alpha v^i - \beta^i$, Eq.~(\ref{eq:continuity2}) can be further
rewritten as
\begin{equation}
\partial_t \left(\sqrt{\gamma} D\right) + \partial_i \left[\sqrt{\gamma} D
  \mathcal{V}^i\right] = 0\,,
\end{equation}
which represents the 3+1 conservative form of the continuity equation in
a generic (curved) spacetime.

Before considering the 3+1 conservative form of the energy-momentum
equations, it is useful to write the energy-momentum tensor in terms of
quantities measured by the Eulerian observer. In particular, we can
define the conserved total energy density $\mathcal{U}$ as the full
projection of the energy-momentum tensor $T^{\mu \nu}$ along the unit
normal $\bm{n}$ to the spatial hypersurface $\Sigma_t$, \ie 
\begin{align}
\mathcal{U} :=& n_\mu n_\nu T^{\mu \nu}  \notag \\
=& \rho h W^2 - p + \frac{1}{2} \left[B^2 \left(1+v^2\right) - \left(B^j
  v_j\right)^2\right]\,.
\end{align}
Similarly, the three-momentum density measured by the Eulerian observer
is defined as the mixed parallel-transverse component of the
energy-momentum tensor
\begin{align}
S_i :=& \gamma^\mu_{\ \,i}  n^\nu T_{\nu \mu} \notag \\
=& \rho h W^2 v_i + B^2 v_i -(B^j v_j)B_i\,.
\end{align}
while the purely spatial part of the energy-momentum tensor is given by
\begin{align}
W^{ij} :=& \gamma^i_{\ \,\mu} \gamma^j_{\ \,\nu} T^{\mu \nu} \notag \\
=& S^i v^j + p_{\rm tot} \gamma^{ij} -\frac{B^i B^j}{W^2} - (B^k v_k) v^i B^j\,.
\end{align}
The corresponding four-dimensional definitions are given respectively by
\begin{equation}
W^{\alpha\beta} := \gamma^{\alpha}_{\ \,\mu} \gamma^{\beta}_{\ \,\nu}
T^{\mu \nu}\,,
\quad {\rm and} \quad S_\alpha := \gamma^\mu_{\ \,\alpha} n^\nu
T_{\nu \mu}\,,
\end{equation}
and allow us to rewrite the energy-momentum tensor in its generic 3+1
decomposition
\begin{equation}
  \label{eq:decompose_Tmunu}
  T^{\mu \nu} = \mathcal{U} n^\mu n^\nu + S^\mu n^\nu + S^\nu n^\mu +
W^{\mu \nu}\,. 
\end{equation}

Next, recalling that the four-divergence of a symmetric rank-2 tensor is
given by
\begin{equation}
  \label{eq:four-divergence}
  \nabla_\mu T^{\mu \nu} = g^{\nu \lambda} \left[ \frac{1}{\sqrt{-g}}
    \partial_\mu \left(\sqrt{-g} T^\mu_{\ \,\lambda}\right) - \frac{1}{2}
    T^{\alpha \beta} \partial_\lambda g_{\alpha \beta} \right]\,,
\end{equation}
we can express the conservation of energy-momentum tensor \eqref{eq:1-2} as
\begin{equation}
 \frac{1}{\sqrt{-g}} \partial_\mu \left(\sqrt{-g} T^\mu_{\ \,\nu}\right)
 =
 \frac{1}{2}
 T^{\mu \lambda} \partial_\nu g_{\mu \lambda}\,,
\end{equation}
and thus, using expression (\ref{eq:decompose_Tmunu}), obtain the 3+1
conservative form of the momentum-conservation equation in a curved
spacetime
\begin{equation}
  \label{eq:momentum_conserv}
  \partial_t \left(\sqrt{\gamma} S_j\right) + \partial_i
  \left[\sqrt{\gamma} \left(\alpha W^i_j - \beta^i S_j\right)\right] =
  \frac{1}{2} \sqrt{-g} T^{\mu \nu} \partial_j T_{\mu \nu}\,.
\end{equation}

In a similar way, we can combine Eqs.~\eqref{eq:1-2} and
\eqref{eq:en_cons} as
\begin{equation}
\nabla_\mu \left(T^{\mu \nu} n_\nu\right) - T^{\mu \nu} \nabla_\mu n_\nu
= 0\,,
\end{equation}
and replacing $T^{\mu \nu}$ with its decomposed
form~\eqref{eq:decompose_Tmunu} so that, after some algebra, we obtain
the 3+1 conservative form of the energy-conservation equation in curved
spacetime
\begin{equation}
  \label{eq:energy_conserv}
  \partial_t \left(\sqrt{\gamma} \mathcal{U}\right) + \partial_i
  \left[\sqrt{\gamma} \left(\alpha S^i - \mathcal{U} \beta^i 
    \right)\right] = -\sqrt{-g} T^{\mu \nu} \nabla_\mu n_\nu \,.
\end{equation}

Equations (\ref{eq:momentum_conserv}) and (\ref{eq:energy_conserv}) both
involve source terms on their right-hand sides. In particular, the source
term of the momentum-conservation equation (\ref{eq:momentum_conserv})
can be written more explicitly as
\begin{align}
\frac{1}{2} \sqrt{-g} T^{\mu \nu} \partial_j g_{\mu \nu} &= \sqrt{-g}
\left( \frac{1}{2} W^{ik} \partial_j \gamma_{ik} + S^\mu n^\nu \partial_j
g_{\mu \nu} + \frac{1}{2} \mathcal{U} n^\mu n^\nu \partial_j g_{\mu \nu}
\right) \notag \\ &= \sqrt{-g} \left( \frac{1}{2} W^{ik} \partial_j
\gamma_{ik} + \frac{1}{\alpha} S_i \partial_j \beta^i - \mathcal{U}
\partial_j \ln{\alpha} \right)\,,
\end{align}
where we have used the following identities
\begin{equation}
\partial_j g_{\mu \nu} = \Gamma^\kappa_{\ \,j \nu} g_{\mu \kappa} +
\Gamma^\kappa_{\ \,j \nu} g_{\kappa \nu}\,,
\end{equation}
and
\begin{equation}
n^\mu \nabla_j n_\mu =0\,.
\end{equation}
where $\Gamma^{\alpha}_{\ \, \beta\gamma}$ are the Christoffel
symbols. Similarly, the source terms of the energy-conservation equation
(\ref{eq:energy_conserv}) can be expressed explicitly as
\begin{equation}
\label{eq:rhs_en_cons}
  - \sqrt{-g} T^{\mu \nu} \nabla_\mu n_\nu = \sqrt{-g} \left(K_{ij} W^{ij}
- S^i \partial_i \ln{\alpha}\right)\,,
\end{equation}
where $K_{\mu \nu}$ is the extrinsic curvature~\cite{Alcubierre:2008,
  Gourgoulhon2012, Rezzolla_book:2013}. Expression \eqref{eq:rhs_en_cons}
can be further simplified if the spacetime is stationary, in which case
the term $\alpha K_{ik} W^{ik}$ reduces to
\begin{equation}
\alpha K_{ik} W^{ik} = \frac{1}{2} W^{ik} \beta^j \partial_j \gamma_{ik}
+ W^j_{\ \,i} \partial_j \beta^i\,.
\end{equation}
We should note that the expressions for the right-hand side of the energy
and momentum conservation equations are due to Ref.~\cite{DelZanna2007}
and do not correspond to the ones originally presented in
Ref.~\cite{Banyuls97}, which expressed the four divergences of the
energy-momentum tensor as
\begin{equation}
\nabla_\mu T^{\mu \nu} = \frac{1}{\sqrt{-g}} \partial_\mu \left(\sqrt{-g}
T^{\mu \nu}\right) + \Gamma^\nu_{\ \,\mu \lambda} T^{\mu \lambda}\,.
\end{equation}
While this expression is mathematically equivalent to
Eq.~\eqref{eq:four-divergence}, it leads to more complex expressions for
the right-hand sides when expressed in a 3+1 decomposition.

The final equation needed to complete the set of GRMHD equations in 
3+1 conservative form is relative to the electromagnetic sector and, in
particular, Faraday's law of induction. Using Eq.~\eqref{eq:Maxwell2} and
the definition of the dual Faraday tensor, it is possible to rewrite
Eq.~\eqref{eq:1-11} in the 3+1 equivalent form
\begin{equation}
  \label{eq:induction}
  \partial_t \sqrt{\gamma} B^j + \partial_i \left[\sqrt{\gamma} B^j
    \left(\alpha v^i - \beta^i\right) - B^i \left(\alpha v^j -
  \beta^j\right)\right] = 0\,,
\end{equation}
or, using the transport velocity, as
\begin{equation}
\label{eq:3p1_indeq}
  \partial_t \sqrt{\gamma} B^j + \partial_i \left[\sqrt{\gamma} \left(B^j
    \mathcal{V}^i - B^i \mathcal{V}^j\right)\right] = 0\,.
\end{equation}
Equation~\eqref{eq:3p1_indeq} is also known as the general-relativistic
induction equation and highlights how the evolution of the magnetic field
is directly related to the curl of the vector product of the
three-velocity and of the magnetic field, as in Newtonian MHD. Once 
the magnetic field is known, the electric field follows trivially from
Eq.~\eqref{eq:ideal_condition}.

At this point, it is possible to collect Eqs.~\eqref{eq:continuity2},
\eqref{eq:momentum_conserv}, \eqref{eq:energy_conserv}, and
\eqref{eq:induction}, so as to build the state vector, the fluxes and the
source vector and hence write the ideal-GRMHD equations in the
conservative Valencia formulation
(\ref{eq:compact-form})~\cite{Anton06}. More specifically, after some
simple algebra, it is not difficult to derive the following expressions
for the state and flux vectors
\begin{equation}
\label{eq:statevec_1}
\bm{U} =
\begin{pmatrix}
D \\
~\\
S_j \\
~\\
\mathcal{U} \\
~\\
B^j \\
\end{pmatrix}\,,
\qquad \qquad
\bm{F}^i =
\begin{pmatrix}
D \mathcal{V}^i \\
~\\
\alpha W^i_{\ \,j} - \beta^i S_j \\
~\\
\alpha S^i - \beta^i \mathcal{U} \\
~\\
B^j \mathcal{V}^i - B^i \mathcal{V}^j \\
\end{pmatrix}\,, 
\end{equation}
while the source vector has components
\begin{equation}
\bm{S} =
\begin{pmatrix}
0 \\
~\\
\frac{1}{2} \alpha W^{ik}\partial_j \gamma_{ik} + S_i \partial_j \beta^i
- \mathcal{U} \partial_j \alpha \\
~\\
\frac{1}{2} W^{ik} \beta^j \partial_j \gamma_{ik} + W^j_i \partial_j
\beta^i -S^j \partial_j \alpha \\
~\\
0 \\
\end{pmatrix}\,.
\label{eq:3+1-GRMHD3}
\end{equation}
We note that since the linear combination of a conserved variable is
still a conserved variable, the state vector of conserved
variables~\eqref{eq:statevec_1} is not unique and different codes
implement different definitions. A particularly common choice is to
replace the conserved total energy density $\mathcal{U}$ with its
equivalent
\begin{equation}
  \label{eq:3+1-GRMHD2}
  \tau := \mathcal{U} - D = \rho W (hW -1) -p + \frac{1}{2} [B^2 (1+v^2) -
  (B_j v^j)^2]\,.
\end{equation}
so that the corresponding state and flux vectors are given by
\begin{equation}
\label{eq:statevec_2}
\bm{U} =
\begin{pmatrix}
D \\
~\\
S_j \\
~\\
\tau \\
~\\
B^j \\
\end{pmatrix}\,,
\qquad \qquad
\bm{F}^i =
\begin{pmatrix}
D \mathcal{V}^i \\
~\\
\alpha W^i_j - \beta^i S_j \\
~\\
\alpha (S^i - v^i D) - \beta^i \tau \\
~\\
B^j \mathcal{V}^i - B^i \mathcal{V}^j \\
\end{pmatrix}\,,
\end{equation}
while the source terms are not changed. Obviously, expressions
\eqref{eq:statevec_1} and \eqref{eq:statevec_2} are mathematically
equivalent, but their numerical implementation has shown that the latter
systematically leads to more accurate evolutions.

A few remarks before concluding this section. First, in stark contrast
with what happens for the conservative formulation of the Newtonian MHD
equations (\ie when $v^2 \ll 1$, $p \ll \rho$, and $E^2 \ll B^2 \ll
\rho$), in GRMHD (but already in general-relativistic hydrodynamics) the
relation between the primitive variables, \ie $\rho$, $v^i$, $\epsilon$
(or $p$), and $B^i$, and the conserved variables, \ie $D$, $S_j$, $\tau$,
and $B^j$, is not analytic. In other words, the calculation of the
primitive variables from the conserved one, \ie what is normally referred
to as the ``primitive recovery'' procedure, cannot be done analytically
but requires the use of a multidimensional root-finding approach. The
latter needs to be implemented at each numerical cell of the
computational domain and can be particularly complex when tabulated
equations of state are employed. Over the years, various algorithms have
been developed to ensure an accurate, efficient, and stable primitive
recovery, aiming at minimizing error accumulation during the matter
evolution. For compactness, we will not discuss here the feature of such
algorithms, but detailed discussions and comparisons of different
algorithms have been studied in Refs.~\cite{Siegel2018a, Kastaun2021,
  Espino2022, Ng2023b} with specific focus on their accuracy and
robustness. Second, the set of evolution
equations (\ref{eq:compact-form}) does not include another important
equation that needs to be solved together such a set and that represents
one of the most important ones in the actual solution of the GRMHD
equations, namely, the divergence-free condition~\eqref{eq:divBeq0}. Also
in this case, we will not discuss here the numerous numerical approaches
that are possible to limit or prevent the growth of the violations of the
divergence-free condition -- from the divergence-cleaning methods, over
to the evolution of the vector potential and up to the sophisticated
constrained-transport methods -- and refer the interested reader to
Ref.~\cite{Olivares2019} for a recent review of the various methods, and
to the following chapters in this book. Finally, in the absence of gravity
and thus in flat spacetimes, the corresponding form of the special
relativistic MHD equations can be obtained trivially after setting
$\alpha=1$, $\beta^i=0$, and $\sqrt{\gamma}=1$. The corresponding set of
equations is widely employed in particle physics to simulate the dynamics
of the collisions of heavy ions and some representative examples can be
found in Refs.~\cite{Inghirami18, Mayer2024}.

\subsection{General-Relativistic Resistive MHD Equations}

As mentioned in the Introduction, the electrical conductivity in
astrophysical plasmas is extremely high and the ideal-MHD condition of
infinite conductivity represents a very good approximation. In this case,
the magnetic flux is conserved and the magnetic field is frozen in the
fluid, being simply advected with it, and the electric field is trivially
obtained from expression~\eqref{eq:ideal_condition}. By construction,
therefore, the solution of the ideal-GRMHD equations neglects any effect
of resistivity on the dynamics. In practice, however, even in
astrophysical plasmas there will be spatial regions with very high
temperatures where the electrical conductivity is finite and the
resistive effects, most notably, the creation of current sheets and the
consequent reconnection, will play a role. Such effects are expected to
take place, for example, during the merger of two magnetized neutron
stars or in accretion disks onto supermassive black holes, and could
provide an important contribution to the energy losses from the system.

In all of these scenarios, the ideal-MHD limit may not be sufficient to
study those physical processes that involve reconnection or the presence
of anisotropic resistivities. This has motivated the derivation of a more
extended set of GRMHD equations that accounts for resistive effects and
that, in practice is augmented by an evolution equation for the electric
field and by an extended Ohm's law. The corresponding system is normally
referred to as the set of general-relativistic resistive MHD (GRRMHD)
equations, and we refer to Ref.~\cite{Palenzuela:2008sf}, where this
system was first presented.

Since the changes with respect to the GRMHD equations take place only in
the electromagnetic sector, the conservation equations of rest-mass
\eqref{eq:continuity2}, momentum \eqref{eq:momentum_conserv}, and energy
\eqref{eq:energy_conserv} remain unchanged. On the other hand, the terms
involved in the basic decomposition of the energy-momentum tensor
\eqref{eq:decompose_Tmunu} are modified because of the explicitly
appearance of the electric fields and take the form
\begin{align}
\mathcal{U} &:= n_\mu n_\nu T^{\mu \nu} = \rho h W^2 -p + \frac{1}{2}
  (E^2 + B^2)\,, \\
S_i &:= \gamma^\mu_{\ \,i} n^\alpha T_{\alpha \mu} = \rho h W^2 v_i +
\sqrt{\gamma} \eta_{ijk} E^k B^k\,, \\
W^{ij} &:= \gamma^i_{\ \,\mu} \gamma^j_{\ \,\nu} T^{\mu \nu} = \rho h W^2
v^i v^j - E^i E^j - B^i B^j + \left[ p + \frac{1}{2} (E^2 + B^2) \right]
\gamma^{ij}\,.
\end{align}

The new expressions for the electromagnetic sector involve the evolution
equation for the magnetic field \eqref{eq:Maxwell2} (Faraday's induction
law), whose modified expression takes into account also the contributions
from the electric fields 
\begin{equation}
\partial_t \sqrt{\gamma} B^j + \partial_i \left[ \sqrt{\gamma} \left(
  \beta^j B^i - \beta^i B^j + \frac{1}{\sqrt{\gamma}} \eta^{ijk} \alpha
  E_k \right) \right] = 0\,. \label{eq:Faraday}
\end{equation}
and the corresponding divergence-free constraint following from the
temporal component of Eq.~(\ref{eq:Maxwell2})
\begin{equation}
  \label{eq:divBeq0}
  \frac{1}{\sqrt{\gamma}} \partial_i \sqrt{\gamma} B^i = 0\,.
\end{equation}
In addition, Ampere's law for the evolution of the electric field follows
from the spatial part of the second couple of Maxwell equations
\eqref{eq:Maxwell2} and is given by
\begin{equation}
\partial_t \sqrt{\gamma} E^j + \partial_i \left[ \sqrt{\gamma} \left(
  \beta^j E^i - \beta^i E^j - \frac{1}{\sqrt{\gamma}} \eta^{ijk} \alpha
  B_k \right) \right] = -\sqrt{\gamma} (\alpha J^i - q\beta^j)\,,
\label{eq:Ampere}
\end{equation}
while the temporal component of \eqref{eq:Maxwell2} expresses the charge
density in terms of the divergence of the electric field, \ie
\begin{equation}
  \frac{1}{\sqrt{\gamma}} \partial_i \left(\sqrt{\gamma} E^i\right) = q\,.
  \label{eq:Gauss}
\end{equation}
Substituting now Eqs.~\eqref{eq:Ohm} and \eqref{eq:Gauss} in Ampere's law
\eqref{eq:Ampere}, we can rewrite it as
\begin{align}
\partial_t &\sqrt{\gamma} E^j + \partial_i \left[ \sqrt{\gamma} \left(
  \beta^j E^i - \beta^i E^j - \frac{1}{\sqrt{\gamma}} \eta^{ijk} \alpha
  B_k \right) \right] \notag \\
&= -\sqrt{\gamma} \frac{\alpha W}{\eta} \left[ E^i +
  \frac{1}{\sqrt{\gamma}} \eta^{ijk} v_j B_k - \left(v_k E^k\right)v^i
  \right] - \left(\alpha v^j -\beta^j\right) \partial_j
\left(\sqrt{\gamma} E^j\right)\,. \label{eq:Ampere2}
\end{align}

At this point, we can finally collect Eqs.~\eqref{eq:continuity2},
\eqref{eq:momentum_conserv}, \eqref{eq:energy_conserv},
\eqref{eq:Faraday}, and \eqref{eq:Ampere}, so as to obtain the
conservative form~\eqref{eq:compact-form} of the GRRMHD equations. The
corresponding augmented state and flux vectors are given by
\begin{equation}
\bm{U} =
\begin{pmatrix}
D \\
~\\
S_j \\
~\\
\mathcal{U} \\
~\\
B^j \\
~\\
E^j \\
\end{pmatrix}\,,
\qquad \qquad
\bm{F}^i =
\begin{pmatrix}
D \mathcal{V}^i \\
~\\
\alpha W^i_j - \beta^i S_j \\
~\\
\alpha S^i - \beta^i \mathcal{U} \\
~\\
\beta^j B^i - \beta^i B^j + \alpha \eta^{ijk} E_k / {\sqrt{\gamma}} \\
~\\
\beta^j E^i - \beta^i E^j - \alpha \eta^{ijk} B_k/ {\sqrt{\gamma}} \\
\end{pmatrix}\,, \label{eq:3+1-GRRMHD}
\end{equation}
while the source terms are 
\begin{equation}
\bm{S} =
\begin{pmatrix}
0 \\
~\\
\frac{1}{2} \alpha W^{ik}\partial_j \gamma_{ik} + S_i \partial_j \beta^i
- \mathcal{U} \partial_j \alpha \\
~\\
\frac{1}{2} W^{ik} \beta^j \partial_j \gamma_{ik} + W^j_i \partial_j
\beta^i -S^j \partial_j \alpha \\
~\\
0 \\
~\\
-\alpha J^j + \beta^j q \\ 
~\\
\end{pmatrix}\,. \label{eq:3+1-GRRMHD2}
\end{equation}

Note that, as for the set of GRMHD equations, also the set of GRRMHD
equations is not closed and requires the specification of an equation of
state. Similarly, a complex primitive-recovery approach is needed to
compute at each numerical cell the primitive variables from the conserved
ones (see, \eg Ref.~\cite{Ripperda2019}). What is new, however, is that
Ohm's law~\eqref{eq:Ohm} does not provide any information on the
properties of the resistivity, which will in general be a function of
space and time, \ie $\eta=\eta(x^i,t)$. In principle, the information
about the properties of the resistivity should follow from microphysical
considerations and hence be calculated under specific physical conditions
(see, \eg \cite{Harutyunyan2018} in the case of neutron-star matter). In
practice, however, given the poor knowledge of the resistive aspects of
astrophysical plasmas, much cruder choices are made that model the
resistivity as a constant or as a simple function of the rest-mass
density (see, \eg \cite{Dionysopoulou:2012pp, Palenzuela2013}). Finally,
it should be mentioned that the solution of the GRRMHD equations is
considerably more challenging that those of the GRMHD equations. For the
former set, in fact, the equations become mixed hyperbolic-parabolic in
Newtonian physics or hyperbolic with stiff relaxation terms in special
relativity. The appearance of stiff terms in the equations follows from
the fact that the diffusive effects take place on timescales that are
intrinsically larger than the dynamical one or, in other words, from the
fact the relaxation terms dominate over the purely hyperbolic ones,
posing severe constraints on the timestep for the evolution (similar
difficulties arise when considering the inclusion of radiative effects;
see, \eg Refs.~\cite{Zanotti2011, Radice2022, Musolino2023}). Overall,
the presence of stiff terms forces the use of specially designed
numerical methods such as the implicit-explicit Runge-Kutta
time-integration schemes or RK-IMEX~\cite{pareschi_2005_ier}, and hence a
significant extension of the numerical infrastructure (see, \eg
Refs.~\cite{Palenzuela2013, Ripperda2019}).

\section{Summary}
\label{sec:1.5}

The equations of MHD represent a very effective tool to describe the
dynamics of astrophysical plasmas and because of this they are employed
in a variety of scenarios in astrophysics and cosmology. When applied in
their GRMHD form to study astrophysical compact objects, they are the
most accurate and powerful tool to explore the properties of compact
black holes and neutron stars in fully dynamical and nonlinear
regimes. In this chapter, we have presented the MHD equations in generic
and curved spacetimes, either in the ideal-MHD limit (GRMHD equations) or
in the presence of resistivity (GRRMHD equations). We have also discussed
the basic aspects of a 3+1 decomposition of spacetime and the importance
of writing the MHD equations in a conservative form.

The purpose of this Chapter -- together with Chapters 2, which covers the
coupling of the MHD equations with the Einstein equations, and of Chapter
3, which reviews the properties of modern equations of state -- is to lay
out the mathematical foundations of the set of equations that will be
employed extensively in the rest of the book and which have a more
applied nature. In particular, after an introduction of some
representative GRMHD codes in Part II, Part III and IV are dedicated to
the study of the dynamics of a variety of astrophysical plasmas, either
when the spacetime is held fixed because the plasmas are not
self-gravitating (Part III) or when the spacetime is dynamically evolved
because the plasmas are self-gravitating (Part IV).


\begin{acknowledgement}
Y.M. is supported by the Shanghai Municipality orientation program of
Basic Research for International Scientists (Grant No. 22JC1410600), the
National Natural Science Foundation of China (Grant No. 12273022), and
the National Key R\&D Program of China (Grant No. 2023YFE0101200).
L.R. is supported in part from the State of Hesse within the Research
Cluster ELEMENTS (Project ID 500/10.006), by the ERC Advanced Grant
``JETSET: Launching, propagation and emission of relativistic jets from
binary mergers and across mass scales'' (Grant No. 884631), and by the
Walter Greiner Gesellschaft zur F\"orderung der physikalischen
Grundlagenforschung e.V. through the Carl W. Fueck Laureatus Chair.

\end{acknowledgement}


\bibliographystyle{plain}
\bibliography{ref}


\end{document}